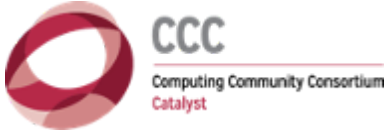

# Artificial Intelligence & Cooperation

*A Computing Community Consortium (CCC) Quadrennial Paper*

*Elisa Bertino (Purdue University), Finale Doshi-Velez (Harvard University), Maria Gini (University of Minnesota), Daniel Lopresti (Lehigh University), and David Parkes (Harvard University)*

The rise of Artificial Intelligence (AI) will bring with it an ever-increasing willingness to cede decision-making to machines. But rather than just giving machines the power to make decisions that affect us, we need ways to work cooperatively with AI systems. There is a vital need for research in "AI and Cooperation" that seeks to understand the ways in which systems of AIs and systems of AIs with people can engender cooperative behavior. Trust in AI is also key: trust that is intrinsic and trust that can only be earned over time. Here we use the term "AI" in its broadest sense, as employed by the recent 20-Year Community Roadmap for AI Research (Gil and Selman, 2019), including but certainly not limited to, recent advances in deep learning.

With success, cooperation between humans and AIs can build society just as human-human cooperation has. Whether coming from an intrinsic willingness to be helpful, or driven through self-interest, human societies have grown strong and the human species has found success through cooperation. We cooperate "in the small" -- as family units, with neighbors, with co-workers, with strangers -- and "in the large" as a global community that seeks cooperative outcomes around questions of commerce, climate change, and disarmament. Cooperation has evolved in nature also, in cells and among animals. While many cases involving cooperation between humans and AIs will be asymmetric, with the human ultimately in control, AI systems are growing so complex that, even today, it is impossible for the human to fully comprehend their reasoning, recommendations, and actions when functioning simply as passive observers.

The research agenda is necessarily broad, involving computer science, economics, psychology, linguistics, law, and philosophy. Indeed, cooperation can mean many different things. The early distributed AI literature studied systems of AIs that all share the same utility function and all want the same things. But we can also consider the economic model of *self-interested, rational agents*, i.e., agents that seek what is best for them individually. Cooperation can also arise here. As is well known from the classical Prisoner's dilemma in game theory, cooperation can also arise in repeated interactions between self-interested agents.

For successful cooperation between people and AI systems, we will need AI systems that understand human preferences, that can model the behavior of others, and that can respond to norms and ethical constructs. We will need AI systems that operate within current laws, institutions and coordination mechanisms and to understand where new kinds of "rules of encounter" will be useful in promoting socially desirable outcomes (Rosenschein and Zlotkin, 1994). We may also need new laws to govern the interactions between AIs and humans. It is also important that AI systems be able to interpret the full range of human communication modalities, and to be able to respond in similar fashion.

The following is a set of illustrative examples of questions that we find important and interesting.

- **Architectures for AI.** Marvin Minsky suggested the "society of minds" paradigm, imagining a loosely coupled, modular reasoning system working together to provide emergent intelligence (Minsky 1988). How can this "AI as system of systems" view be realized? What does it take for robust AI to emerge from modular components -- how to promote diverse models, how to arbitrate, how to learn complementary skills? For example, a car will have multiple inputs and multiple reasoning systems, so you need a system point of view; can you have an independent AI system that coordinates what all the various components do? In robotics, you have agents controlling different parts of the robotic system, one controls the arm, one the leg, etc.; can this be usefully modeled as a cooperative AI system? At home, you have an AI that controls your house, another your car; can you have an AI that has an overview of both, providing coordination while at the same time adhering to your own stated policies and preferences?
- **Collaborative Human-AI Systems**. How do we build AI systems that work effectively with people across different contexts -- complementing each other, and with AI systems that move beyond explanations and support agency on the part of users? In regard to explanations, how should these vary according to time constraints, contexts and tasks? How can we design AI systems to form a part of a team and work collaboratively with others (Grosz and Kraus 1996)? Can we understand how to transfer cooperative modalities from one setting to another? Which kinds of human decision-making processes can benefit from "AI supervision," and which kinds of supervision should be implemented? What is the role of *modeling* people, so that AI systems can learn to be complementary -- presenting the right information at the right time and in the right way, while intervening when helpful?
- **Markets, Mechanism Design, and Economic Viewpoints**. Mechanism design is the field of microeconomics that studies incentive alignment and looks for ways to promote beneficial system-wide outcomes in economic systems, and to do so despite self-interest and private information. It can be viewed as "inverse game theory" in that it looks to design rules of engagement such that the implied equilibrium is beneficial (e.g., allocating resources to maximize value, finding the best possible compromise -- a Pareto optimal compromise -- in a shared decision problem, or assigning tasks to those with the most appropriate skills). What role can mechanism design play in the development of cooperative AI? Classical theory considers static, centralized, and direct mechanisms -- the rules to govern AI systems will need to be richer, and embrace dynamics, decentralization, and partial information about preferences (Parkes and Wellman 2015). Market systems also play an important role in human society in allocating capital, resources, and labor. What kinds of market systems will be effective in promoting good outcomes in AI economies?

- **Understanding Human Preferences.** There is a difference between my "revealed preferences" (I eat a candy bar) and my "latent preferences" (I think today that I should go to the gym tomorrow). Because of decision biases, present-bias as an example, my revealed preference may not reflect my latent preferences. As a result, an AI system that passively observes behavior and infers our preferences, in the spirit of inverse reinforcement learning, will learn our revealed preferences but not our latent preferences. An AI system that acts on our behalf would then promote "candy bar" style decisions rather than "going to the gym" style decisions. Is this what it should mean for the AI system to cooperate with us, or not? If not, then what is the role for preference elicitation as a way to actively elicit our "should" preferences?
- **Control and Related Concerns.** In the drive towards *helpful AI* (Hadfield-Menell et al. 2016), there are concerns about loss of control: if we cede decisions to AI, and then AI starts to make decisions, and we stop paying attention, and then we cede more decisions to AI, and so forth, then which decisions are we left with making, and how do we ensure that AI systems continue to change as *human* societal preferences change (even knowing this requires feedback from users). Our societies are extremely complex and we evolve in our thinking about what is right and what is wrong and in response to injustice and inequity. How can cooperative AI be designed in a way that continues to provide people with the right control, awareness, transparency and ability to govern? Do we have a future of "AI fitting to people" or "people fitting to AI," and might an increased prevalence of automated decisions come to change our preferences over time? What about concerns around *manipulation* -- the same ability to predict human behavior that can be used in allowing AI to be cooperative with people can be used to allow AI systems to be designed to promote decisions that favor one party over another in ways that may be unfair. Another important concern relates to *fairness*, and to the potential for *bias* in badly-conceived AI systems, and to the need to ensure that AI systems cooperate equally well across all parts of society.

The AI roadmap (Gil and Selman, 2019) also describes themes that fit with this research agenda on AI and cooperation. Section 3.2 is titled "A Research Roadmap for Meaningful Interaction" and includes the following two directions (presented here in summary):

- **Enabling Collaborative Interaction**: Interactions with today's AI systems lack the elegance and ease of typical interactions between people. Building AI systems that interact with humans as fluently as people work together will require solving challenges ranging from explaining their reasoning and modeling humans' mental states to understanding social norms and supporting complex teamwork.
- **Making AI Systems Trustworthy**: Based on their understanding of an AI system, users should be able to assess and improve any undesirable behaviors. AI systems should have provisions to avoid undesirable persuasion and manipulation. Appropriate mechanisms need to be developed to enable AI systems to act responsibly and to engender and uphold people's trust.

At a time when we have increasing success with narrow AI -- with the application of AI to particular problems such as vision, natural language, or planning -- it is time to embrace a broader agenda around what it means to bring together AI systems with people in a way that engenders cooperative behavior. This shift in perspective will be vital in advancing AI to a level where it can serve as a true and trustworthy partner to humans. We recommend a federally funded research program on "AI and Cooperation" that is grounded in technical research (e.g., embracing machine learning and probabilistic inference, automated reasoning about preferences, planning, multi-agent learning, and game theory), including theory and experimentation, and also deliberately multi-disciplinary, seeking to bring together viewpoints from fields such as psychology, linguistics, law, and philosophy.

*This white paper is part of a series of papers compiled every four years by the CCC Council and members of the computing research community to inform policymakers, community members and the public on important research opportunities in areas of national priority. The topics chosen represent areas of pressing national need spanning various subdisciplines of the computing research field. The white papers attempt to portray a comprehensive picture of the computing research field detailing potential research directions, challenges and recommendations.*

*This material is based upon work supported by the National Science Foundation under Grant No. 1734706. Any opinions, findings, and conclusions or recommendations expressed in this material are those of the authors and do not necessarily reflect the views of the National Science Foundation.*

*For citation use: Bertino E., Finale D., Gini M, Lopresti D. & Parkes D. (2020) Artificial Intelligence & Cooperation.*
*https://cra.org/ccc/resources/ccc-led-whitepapers/#2020-quadrennial-papers*